\documentclass{llncs}

\usepackage{t1enc} 
\usepackage{amsmath}
\usepackage{latexsym}
\usepackage{theorem}
\usepackage{amssymb}
\usepackage{url}

\newcommand{\comment}[1]{}

\newcommand{\hsp}[1][3mm]{\hspace*{#1}}
\newcommand{\vsp}[1][3mm]{\vspace*{#1}}
\newcommand{\moins}{\setminus}
\newcommand{\vide}{\emptyset}
\newcommand{\ie}{{\em i.e.} }
\newcommand{\eg}{{\em e.g.} }
\newcommand{\etal}{{\em et al} }

\newcommand{\bN}{\mb{N}}
\newcommand{\bZ}{\mb{Z}}
\newcommand{\bQ}{\mb{Q}}

\newcommand{\xu}{\{x\to u\}}
\newcommand{\xv}{\{x\to v\}}

\newcommand{\dom}{\mr{dom}}

\newcommand{\FV}{\mr{FV}}

\renewcommand{\a}{\rightarrow}
\newcommand{\A}{\Rightarrow}

\renewcommand{\AA}{\Leftrightarrow}

\newcommand{\ad}{\downarrow}

\renewcommand{\to}{\mapsto}

\newcommand{\al}[2][]{~_{#2}^{#1}\!\!\leftarrow}
\newcommand{\als}[1]{\al[*]{#1}}

\newcommand{\ex}{\exists}
\newcommand{\all}{\forall}
\newcommand{\ou}{\vee}

\newcommand{\biget}{\bigwedge}
\newcommand{\et}{\wedge}

\newcommand{\st}{\star}
\newcommand{\B}{\Box} 
\renewcommand{\th}{\vdash}

\newcommand{\sle}{\subseteq}
\newcommand{\sge}{\supseteq}

\newcommand{\qle}{\sqsubseteq}

\newcommand{\lex}{_\mr{lex}}

\renewcommand{\b}{\beta}
\newcommand{\g}{\gamma}
\newcommand{\G}{\Gamma}

\renewcommand{\t}{\theta}

\newcommand{\la}{\lambda}

\renewcommand{\r}{\rho}
\newcommand{\s}{\sigma}
\renewcommand{\S}{\Sigma}
\newcommand{\up}{\upsilon}

\newcommand{\vphi}{\varphi}
\newcommand{\w}{\omega}

\newcommand{\mi}{\mathit}
\newcommand{\mc}{\mathcal}
\newcommand{\mt}{\mathtt}
\newcommand{\mr}{\mathrm}
\newcommand{\mb}{\mathbb}
\newcommand{\mg}{\mathbf}

\newcommand{\cA}{\mc{A}}

\newcommand{\cC}{\mc{C}}
\newcommand{\cD}{\mc{D}}
\newcommand{\cE}{\mc{E}}
\newcommand{\cF}{\mc{F}}

\newcommand{\cR}{\mc{R}}
\newcommand{\cS}{\mc{S}}
\newcommand{\cT}{\mc{T}}

\newcommand{\cV}{\mc{V}}

\newcommand{\cX}{\mc{X}}
\newcommand{\cY}{\mc{Y}}
\newcommand{\cZ}{\mc{Z}}

\newcommand{\va}{{\vec{a}}}

\newcommand{\vl}{{\vec{l}}}

\newcommand{\vt}{{\vec{t}}}
\newcommand{\vu}{{\vec{u}}}
\newcommand{\vv}{{\vec{v}}}

\newcommand{\vx}{{\vec{x}}}

\newcommand{\vT}{{\vec{T}}}

\newenvironment{rul}
  {$\begin{array}{rcl}}
  {\end{array}$}
\newenvironment{rulc}
  {\begin{center}\begin{rul}}
  {\end{rul}\end{center}}

\newenvironment{rew}[1][~~\a~~]
  {$\begin{array}{r@{#1}l}}
  {\end{array}$}
\newenvironment{rewc}[1][~~\a~~]
  {\begin{center}\begin{rew}[#1]}
  {\end{rew}\end{center}}

  {\end{array}$\end{center}}

{\theorembodyfont{\rmfamily} 

}

\newenvironment{lstgeneric}[2]
  {\begin{list}{#1}{\topsep=.5mm\itemsep=.5mm\parsep=0mm%
    \itemindent=-3ex\labelsep=1ex\labelwidth=0ex #2}}
  {\end{list}}

\newenvironment{lst}[1]
  {\begin{lstgeneric}{#1}{\itemindent=-1ex}}
  {\end{lstgeneric}}

\newenvironment{enumi}[1]
  {\begin{lstgeneric}{}{\usecounter{enumi}\leftmargin=7mm%
    }}
  {\end{lstgeneric}}

\newcommand{\ab}{\a_\b}
\newcommand{\ar}{\a_\cR}

\newcommand{\tf}{{\tau_f}}

\newcommand{\tC}{{\tau_C}}

\newcommand{\CF}{\cC\cF}
\newcommand{\CFB}{\CF^\B}

\newcommand{\ry}{\rho_{\!_\cY}}

\newcommand{\bS}{\mb{S}}

\newcommand{\nf}{\!\!\ad}

\newcommand{\tha}[1][\cY]{\th_\mr{\!\!a}^{\!\!^{#1}}}

\newcommand{\les}{\le_\cA}
\newcommand{\eqs}{\simeq_\cA}

\newcommand{\emat}{\end{array}\right)}
\newcommand{\matc}{\left(\begin{array}{@{}r@{}}}
\newcommand{\matcc}{\left(\begin{array}{@{}r@{\,}r@{}}}
\newcommand{\matccc}{\left(\begin{array}{@{}r@{\,}r@{\,}r@{}}}

\newcommand{\si}{\mi{if}}
\newcommand{\alors}{\mi{then}}
\newcommand{\sinon}{\mi{else}}
\newcommand{\vrai}{\mi{true}}
\newcommand{\faux}{\mi{false}}
\newcommand{\dans}{\mi{in}}

\newcommand{\asort}[1][s]{\mt{#1}}

\begin{document}


\title{Decidability of Type-checking in the Calculus of Algebraic Constructions with Size Annotations}

\author{Fr\'ed\'eric Blanqui}

\institute{Laboratoire Lorrain de Recherche en Informatique et Automatique (LORIA)\\
Institut National de Recherche en Informatique et Automatique (INRIA)\\
615 rue du Jardin Botanique, BP 101, 54602 Villers-l\`es-Nancy, France\\
\email{blanqui@loria.fr}}

\maketitle
\thispagestyle{empty}

\begin{abstract}
Since Val Tannen's pioneering work on the combination of simply-typed
$\la$-calculus and first-order rewriting \cite{breazu88lics}, many
authors have contributed to this subject by extending it to richer
typed $\la$-calculi and rewriting paradigms, culminating in the
Calculus of Algebraic Constructions. These works provide theoretical
foundations for type-theoretic proof assistants where functions and
predicates are defined by oriented higher-order equations. This kind
of definitions subsumes usual inductive definitions, is easier to
write and provides more automation.

\hsp On the other hand, checking that such user-defined rewrite rules,
when combined with $\b$-reduction, are strongly normalizing and
confluent, and preserve the decidability of type-checking, is more
difficult. Most termination criteria rely on the term structure. In a
previous work, we extended to dependent types and higher-order
rewriting, the notion of ``sized types'' studied by several authors in
the simpler framework of ML-like languages, and proved that it
preserves strong normalization.

\hsp The main contribution of the present paper is twofold. First, we prove
that, in the Calculus of Algebraic Constructions with size
annotations, the problems of type inference and type-checking are
decidable, provided that the sets of constraints generated by size
annotations are satisfiable and admit most general solutions. Second,
we prove the latter properties for a size algebra rich enough for
capturing usual induction-based definitions and much more.
\end{abstract}


\section{Introduction}

The notion of ``sized type'' was first introduced in
\cite{hughes96popl} and further studied by several authors
\cite{gimenez98icalp,barthe04mscs,abel04ita,xi02hosc} as a tool for proving the
termination of ML-like function definitions. It is based on the
semantics of inductive types as fixpoints of monotone operators,
reachable by transfinite iteration. For instance, natural numbers are
the limit of $(S_i)_{i<\w}$, where $S_i$ is the set of natural numbers
smaller than $i$ (inductive types with constructors having functional
arguments require ordinals bigger than $\w$). The idea is then to
reflect this in the syntax by adding size annotations on types
indicating in which subset $S_i$ a term is. For instance, subtraction
on natural numbers can be assigned the type $-:nat^\alpha\A nat^\b\A
nat^\alpha$, where $\alpha$ and $\b$ are implicitly universally
quantified, meaning that the size of its output is not bigger than the
size of its first argument. Then, one can ensure termination by
restricting recursive calls to arguments whose size -- by typing -- is
smaller. For instance, the following ML-like definition of
$\lceil\frac{x}{y+1}\rceil$:

\begin{center}
\begin{minipage}{5cm}
\small
\begin{verbatim}
letrec div x y = match x with
  | O -> O
  | S x' -> S (div (x' - y) y)
\end{verbatim}
\end{minipage}
\end{center}

\noindent
is terminating since, if $x$ is of size at most $\alpha$ and $y$ is of
size at most $\b$, then $x'$ is of size at most $\alpha-1$ and
$(x'-y)$ is of size at most $\alpha-1<\alpha$.

The Calculus of Constructions (CC) \cite{coquand88ic} is a powerful
type system with polymorphic and dependent types, allowing to encode
higher-order logic. The Calculus of Algebraic Constructions (CAC)
\cite{blanqui05mscs} is an extension of CC where functions are defined
by higher-order rewrite rules. As shown in \cite{blanqui05fi}, it
subsumes the Calculus of Inductive Constructions (CIC)
\cite{coquand88colog} implemented in the Coq proof assistant
\cite{coqv80}, where functions are defined by induction. Using
rule-based definitions has numerous advantages over induction-based
definitions: definitions are easier (\eg Ackermann's function), more
propositions can be proved equivalent automatically, one can add
simplification rules like associativity or using rewriting modulo AC
\cite{blanqui03rta}, etc. For proving that user-defined rules
terminate when combined with $\b$-reduction, \cite{blanqui05mscs}
essentially checks that recursive calls are made on structurally
smaller arguments.

In \cite{blanqui04rta}, we extended the notion of sized type to CAC,
giving the Calculus of Algebraic Constructions with Size Annotations
(CACSA). We proved that, when combined with $\b$-reduction,
user-defined rules terminate essentially if recursive calls are made
on arguments whose size -- by typing -- is strictly smaller, by
possibly using lexicographic and multiset comparisons. Hence, the
following rule-based definition of $\lceil\frac{x}{y+1}\rceil$:

{\small\begin{rewc}
0~/~y & 0\\
(s~x)~/~y & s~((x~-~y)~/~y)\\
\end{rewc}}

\noindent
is terminating since, in the last rule, if $x$ is of size at most
$\alpha$ and $y$ is of size at most $\b$, then $(s~x)$ is of size at
most $\alpha+1$ and $(x-y)$ is of size at most $\alpha<\alpha+1$. Note
that this rewrite system cannot be proved terminating by criteria only
based on the term structure, like RPO or its extensions to
higher-order terms \cite{jouannaud99lics,walukiewicz03jfp}. Note also
that, if a term $t$ is structurally smaller than a term $u$, then the
size of $t$ is smaller than the size of $u$. Therefore, CACSA proves
the termination of any induction-based definition like CIC/Coq, but
also definitions like the previous one. To our knowledge, this is the
most powerful termination criterion for functions with polymorphic and
dependent types like in Coq. The reader can find other convincing
examples in \cite{blanqui04rta}.

However, \cite{blanqui04rta} left an important question open. For the
termination criterion to work, we need to make sure that size
annotations assigned to function symbols are valid. For instance, if
subtraction is assigned the type $-:nat^\alpha\A nat^\b\A nat^\alpha$,
then we must make sure that the definition of $-$ indeed outputs a
term whose size is not greater than the size of its first
argument. This amounts to check that, for every rule in the definition
of $-$, the size of the right hand-side is not greater than the size
of the left hand-side. This can be easily verified by hand if, for
instance, the definition of $-$ is as follows:

{\small\begin{rewc}
0~-~x & 0\\
x~-~0 & x\\
(s~x)~-~(s~y) & x~-~y\\
\end{rewc}}

The purpose of the present work is to prove that this can be done
automatically, by inferring the size of both the left and right
hand-sides, and checking that the former is smaller than the latter.

\begin{figure}[ht]
\caption{Insertion sort on polymorphic and dependent lists\label{fig-sort}}
\begin{rulc}
nil &:& (A:\st)list^\alpha A~0\\ cons &:& {(A:\st)}A\A
{(n:nat)}{list^\alpha A~n}\A{list^{s\alpha} A~(sn)}\\
\si\_\dans\_\alors\_\sinon &:& bool\A{(A:\st)}A\A A\A A\\
insert &:& {(A:\st)}{(\le:A\A A\A bool)}
A\A{(n:nat)}{list^\alpha A~n}\A {list^{s\alpha} A~(sn)}\\
sort &:& {(A:\st)}{(\le:A\A A\A bool)}
{(n:nat)}{list^\alpha A~n}\A {list^\alpha A~n}\\
\end{rulc}

\begin{rulc}
\si~\vrai~\dans~A~\alors~u~\sinon~v &~~\a~~& u\\
\si~\faux~\dans~A~\alors~u~\sinon~v &~~\a~~& v\\

insert~A~\le~x~\_~(nil~\_) &~~\a~~& cons~A~x~0~(nil~A)\\
insert~A~\le~x~\_~(cons~\_~y~n~l) &~~\a~~& \si~x\le y~\dans~list~A~(s~(s~n))\\
&& \alors~cons~A~x~(s~n)~(cons~A~y~n~l)\\
&& \sinon~cons~A~y~(s~n)~(insert~A~\le~x~n~l)\\

sort~A~\le~\_~(nil~\_) &~~\a~~& nil~A\\
sort~A~\le~\_~(cons~\_~x~n~l) &~~\a~~& insert~A~\le~x~n~(sort~A~\le~n~l)\\
\end{rulc}
\end{figure}

We now give an example with dependent and polymorphic types. Let
$\st$ be the sort of types and $list:\st\A nat\A\st$ be the type of
polymorphic lists of fixed length whose constructors are $nil$ and
$cons$. Without ambiguity, $s$ is used for the successor function both
on terms and on size expressions. The functions $insert$ and $sort$
defined in Figure \ref{fig-sort} have size annotations satisfying our
termination criterion. The point is that $sort$ preserves the size of
its list argument and thus can be safely used in recursive
calls. Checking this automatically is the goal of this work.

An important point is that the ordering naturally associated with size
annotations implies some subtyping relation on types. The combination
of subtyping and dependent types (without rewriting) is a difficult
subject which has been studied by Chen \cite{chen98thesis}. We reused
many ideas and techniques of his work for designing CACSA and proving
important properties like $\b$-subject reduction (preservation of
typing under $\b$-reduction) \cite{blanqui04rtafull}.

Another important point is related to the meaning of type
inference. In ML, type inference means computing a type of a term in
which the types of free and bound variables, and function symbols
({\tt letrec}'s in ML), are unknown. In other words, it consists in
finding a simple type for a pure $\la$-term. Here, type inference
means computing a CACSA type, hence dependent and polymorphic (CACSA
contains Girard's system F), of a term in which the types and size
annotations of free and bound variables, and function symbols, are
known. In dependent type theories, this kind of type inference is
necessary for type-checking \cite{coquand91book}. In other words, we
do not try to infer relations between the sizes of the arguments of a
function and the size of its output like in
\cite{chin01hosc,barthe05tlca}. We try to check that,
with the annotated types declared by the user for its function
symbols, rules satisfy the termination criterion described in
\cite{blanqui04rta}.

Moreover, in ML, type inference amounts to solve equality constraints
in the type algebra. Here, type inference amounts to solve equality
and ordering constraints in the size algebra. The point is that the
ordering on size expressions is not anti-symmetric: it is a
quasi-ordering. Thus, we have a combination of unification and
symbolic quasi-ordering constraint solving.

Finally, because of the combination of subtyping and dependent typing,
the decidability of type-checking requires the existence of minimal
types \cite{chen98thesis}. Thus, we must also prove that a satisfiable
set of equality and ordering constraints has a smallest solution,
which is not the case in general. This is in contrast with
non-dependently typed frameworks.

{\bf Outline.} In Section \ref{sec-cac}, we define terms and types,
and study some properties of the size ordering. In Section
\ref{sec-dec}, we give a general type inference algorithm and prove
its correctness and completeness under general assumptions on
constraint solving. Finally, in Section \ref{sec-cons}, we prove that
these assumptions are fulfilled for the size algebra introduced in
\cite{barthe04mscs} which, although simple, is rich enough for
capturing usual inductive definitions and much more, as shown by the
first example above. Missing proofs are given in
\cite{blanqui05cslfull}.


\section{Terms and types}
\label{sec-cac}

{\bf Size algebra.} Inductive types are annotated by {\em size
expressions} from the following algebra $\cA$:

\begin{center}
$a ::= \alpha ~|~ sa ~|~ \infty$
\end{center}

\noindent
where $\alpha\in\cZ$ is a {\em size variable}. The set $\cA$ is
equipped with the quasi-ordering $\les$ defined in Figure
\ref{fig-sord}. Let $\simeq_\cA\,=\,{\le_\cA\cap\ge_\cA}$ be its
associated equivalence.

Let $\vphi,\psi,\rho,\ldots$ denote size substitutions, \ie functions
from $\cZ$ to $\cA$. One can easily check that $\les$ is stable by
substitution: if $a\les b$ then $a\vphi\les b\vphi$. We extend $\les$
to substitutions: $\vphi\les\psi$ iff, for all $\alpha\in\cZ$,
$\alpha\vphi\les\alpha\psi$.

We also extend the notion of ``more general substitution'' from
unification theory as follows: $\vphi$ is {\em more general than}
$\psi$, written $\vphi\qle\psi$, iff there is $\vphi'$ such that
$\vphi\vphi'\les\psi$.


\begin{figure}[ht]
\caption{Ordering on size expressions\label{fig-sord}}
\centering\vsp[3mm]
(refl)~ $a\les a$\hsp[5mm]
(trans)~ $\cfrac{a\les b\quad b\les c}{a\les c}$\\[3mm]
(mon)~ $\cfrac{a\les b}{sa\les sb}$\hsp[5mm]
(succ)~ $\cfrac{a\les b}{a\les sb}$\hsp[5mm]
(infty)~ $a\les\infty$
\end{figure}


{\bf Terms.} We assume the reader familiar with typed $\la$-calculi
\cite{barendregt92book} and rewriting \cite{dershowitz90book}. Details
on CAC(SA) can be found in \cite{blanqui05mscs,blanqui04rta}. We
assume given a set $\cS=\{\st,\B\}$ of {\em sorts} ($\st$ is the sort
of types and propositions; $\B$ is the sort of predicate types), a set
$\cF$ of function or predicate {\em symbols}, a set $\CFB\sle\cF$ of
{\em constant predicate symbols}, and an infinite set $\cX$ of {\em
term variables}. The set $\cT$ of terms is:

\begin{center}
$t ::= \asort ~|~ x ~|~ C^a ~|~ f ~|~ [x:t]t ~|~ (x:t)t ~|~ tt$
\end{center}

\noindent
where $\asort\in\cS$, $x\in\cX$, $C\in\CFB$, $a\in\cA$ and
$f\in\cF\moins\CFB$. A term $[x:t]u$ is an {\em abstraction}. A term
$(x:T)U$ is a {\em dependent product}, simply written $T\A U$ when $x$
does not occur in $U$. Let $\vt$ denote a sequence of terms
$t_1,\ldots,t_n$ of length $|\vt|=n$.

Every term variable $x$ is equipped with a sort $\asort_x$ and, as
usual, terms equivalent modulo sort-preserving renaming of bound
variables are identified. Let $\cV(t)$ be the set of size variables in
$t$, and $\FV(t)$ be the set of term variables free in $t$. Let
$\t,\s,\ldots$ denote term substitutions,
\ie functions from $\cX$ to $\cT$. For our previous examples, we have
$\CFB=\{nat,list,bool\}$ and
$\cF=\CFB\cup\{0,s,/,nil,cons,insert,sort\}$.


{\bf Rewriting.} Terms only built from variables and symbol
applications $f\vt$ are said to be {\em algebraic}. We assume given a
set $\cR$ of {\em rewrite rules} $l\a r$ such that $l$ is algebraic,
$l=f\vl$ with $f\notin\CFB$ and $\FV(r)\sle\FV(l)$. Note that, while
left hand-sides are algebraic and thus require syntactic matching
only, right hand-sides may have abstractions and
products. $\b$-reduction and rewriting are defined as usual:
$C[[x:T]u~v]\ab C[u\xv]$ and $C[l\s]\ar C[r\s]$ if $l\a r\in\cR$. Let
${\a}={\ab\cup\ar}$ and $\a^*$ be its reflexive and transitive
closure. Let $t\ad u$ iff there exists $v$ such that $t\a^* v\als{}
u$.


{\bf Typing.} We assume that every symbol $f$ is equipped with a sort
$\asort_f$ and a type $\tf=(\vx:\vT)U$ such that, for all rules
$f\vl\a r\in\cR$, $|\vl|\le|\vT|$ ($f$ is not applied to more
arguments than the number of arguments given by $\tf$). Let
$\cF^{\asort}$ (resp. $\cX^{\asort}$) be the set of symbols
(resp. variables) of sort $\asort$. As usual, we distinguish the
following classes of terms where $t$ is any term:

\begin{lst}{--}
\item objects: $o ::= x\in\cX^\st ~|~ f\in\cF^\st ~|~ [x:t]o ~|~ ot$
\item predicates: $p ::= x\in\cX^\B ~|~ C^a\in\CFB ~|~ f\in{\cF^\B\moins\CFB}
~|~ [x:t]p ~|~ (x:t)p ~|~ pt$
\item kinds: $K ::= \st ~|~ (x:t)K$
\end{lst}

Examples of objects are the constructors of inductive types
$0,s,nil,cons,\ldots$ and the function symbols
$-,/,insert,sort,\ldots$. Their types are predicates: inductive types
$bool,nat,list,\ldots$, logical connectors $\et,\ou,\ldots$, universal
quantifications $(x:T)U,\ldots$ The types of predicates are kinds:
$\st$ for types like $bool$ or $nat$, $\st\A nat\A\st$ for $list$,
\ldots

An {\em environment} $\G$ is a sequence of variable-term pairs. An
environment is {\em valid} if a term is typable in it. The typing
rules of CACSA are given in Figure \ref{fig-typ} and its subtyping
rules in Figure \ref{fig-sub}. In (symb), $\vphi$ is an arbitrary size
substitution. This reflects the fact that, in type declarations, size
variables are implicitly universally quantified, like in ML. In
contrast with \cite{chen98thesis}, subtyping uses no sorting
judgment. This simplification is justified in \cite{blanqui04rtafull}.

In comparison with \cite{blanqui04rtafull}, we added the side
condition $\cV(\vt)=\vide$ in (size). It does not affect the
properties proved in \cite{blanqui04rtafull} and ensures that the size
ordering is compatible with subtyping (Lemma \ref{lem-le-sub}). By the
way, one could think of taking the more general rule $C^a\vt\le
C^b\vu$ with $\vt\eqs\vu$. This would eliminate the need for equality
constraints and thus simplify a little bit the constraint solving
procedure. More generally, one could think in taking into account the
monotony of type constructors by having, for instance, $list~nat^a\le
list~nat^b$ whenever $a\les b$. This requires extensions to Chen's
work \cite{chen98thesis} and proofs of many non trivial properties of
\cite{blanqui04rtafull} again, like Theorem \ref{thm-sub-s} below or
subject reduction for $\b$.


\begin{figure}[ht]
\centering
\caption{Subtyping rules\label{fig-sub}}
\begin{tabular}{c}
\\(refl)\quad $T\le T$\quad\quad

(size)\quad $C^a\vt\le C^b\vt$\quad ($C\in\CFB$, $a\le_\cA b$,
$\cV(\vt)=\vide$)\\

\\(prod)\quad $\cfrac{U'\le U \quad V\le V'}{(x:U)V\le (x:U')V'}$\quad\quad

(conv)\quad $\cfrac{T'\le U'}{T\le U}$\quad
($T\ad T'$, $U'\ad U$)\\

\\(trans)\quad $\cfrac{T\le U\quad U\le V}{T\le V}$\\
\end{tabular}
\end{figure}


\begin{figure}[ht]
\centering
\caption{Typing rules\label{fig-typ}}
\begin{tabular}{c}
\\(ax)\quad $\th\st:\B$\quad\quad

(prod)\quad $\cfrac{\G\th U:\asort \quad \G,x:U \th V:\asort'}
{\G\th (x:U)V:\asort'}$\\

\\(size)\quad $\cfrac{\th\tC:\B}{\th C^a:\tC}$\quad
($C\in\CFB$, $a\in\cA$)\quad\quad

(symb)\quad $\cfrac{\th\tf:\asort_f}{\th f:\tf\vphi}$\quad
($f\notin\CFB$)\\

\\(var)~$\cfrac{\G\th T:\asort_x}{\G,x:T\th x:T}$~$(x\!\notin\!\dom(\G))$\quad\quad

(weak)~$\cfrac{\G\th t:T \quad \G\th U:\asort_x}{\G,x:U\th t:T}$~$(x\!\notin\!\dom(\G))$\\

\\(abs)\quad $\cfrac{\G,x:U \th v:V \quad \G\th (x:U)V:\asort}
{\G\th [x:U]v:(x:U)V}$\quad\quad

(app)\quad $\cfrac{\G\th t:(x:U)V \quad \G\th u:U}
{\G\th tu:V\xu}$\\

\\(sub)\quad $\cfrac{\G\th t:T \quad \G\th T':\asort}{\G\th t:T'}$\quad
($T\le T'$)\\[3mm]
\end{tabular}
\end{figure}


{\bf $\infty$-Terms.} An {\em $\infty$-term} is a term whose only size
annotations are $\infty$. In particular, it has no size variable. An
{\em $\infty$-environment} is an environment made of
$\infty$-terms. This class of terms is isomorphic to the class of
(unannotated) CAC terms. Our goal is to be able to infer annotated
types for these terms, by using the size annotations given in the type
declarations of constructors and function symbols
$0,s,/,nil,cons,insert,sort,\ldots$

Since size variables are intended to occur in object type declarations
only, and since we do not want matching to depend on size annotations,
we assume that rules and type declarations of predicate symbols
$nat,bool,list,\ldots$ are made of $\infty$-terms. As a consequence,
we have:

\begin{lemma}
\label{lem-size-sub}
\begin{lst}{--}
\item If $t\a_\cR t'$ then, for all $\vphi$, $t\vphi\a_\cR t'\vphi$.
\item If $\G\th t:T$ then, for all $\vphi$, $\G\vphi\th t\vphi:T\vphi$.
\end{lst}
\end{lemma}


\noindent
{\bf We make three important assumptions:}

\begin{enumi}{}
\item $\cR$ preserves typing: for all $l\a r\in\cR$, $\G$, $T$ and $\s$,
if $\G\th l\s:T$ then $\G\th r\s:T$. It is generally not too difficult
to check this by hand. However, as already mentioned in
\cite{blanqui04rta}, finding sufficient conditions for this to
hold in general does not seem trivial.

\item $\b\cup\cR$ is confluent. This is for instance the case if $\cR$
is confluent and left-linear \cite{muller92ipl}, or if $\b\cup\cR$ is
terminating and $\cR$ is locally confluent.

\item $\b\cup\cR$ is terminating. In \cite{blanqui04rta}, it is proved that
$\b\cup\cR$ is terminating essentially if, in every rule $f\vl\a
r\in\cR$, recursive calls in $r$ are made on terms whose size -- by
typing -- are smaller than $\vl$, by using lexicographic and multiset
comparisons. Note that, with type-level rewriting, confluence is
necessary for proving termination
\cite{blanqui05mscs}.
\end{enumi}

\noindent
{\bf Important remark.} One may think that there is some vicious
circle here: we assume the termination for proving the decidability of
type-checking, while type-checking is used for proving termination!
The point is that termination checks are done incrementally. At the
beginning, we can check that some set of rewrite rules $\cR_1$ is
terminating in the system with $\b$ only. Indeed, we do not need to
use $\cR_1$ in the type conversion rule (conv) for typing the terms of
$\cR_1$. Then, we can check in $\b\cup\cR_1$ that some new set of
rules $\cR_2$ is terminating, and so on\ldots


Various properties of CACSA have already been studied in
\cite{blanqui04rtafull}. We refer the reader to this paper if
necessary. For the moment, we just mention two important and non
trivial properties based on Chen's work on subtyping with dependent
types \cite{chen98thesis}: subject reduction for $\b$ and transitivity
elimination:

\begin{theorem}[\cite{blanqui04rtafull}]
\label{thm-sub-s}
$T\le U$ iff ${T\nf}\le_s {U\nf}$, where $\le_s$ is the restriction of
$\le$ to (refl), (size) and (prod).
\end{theorem}


We now give some properties of the size and substitution
orderings. Let $\a_\cA$ be the confluent and terminating relation on
$\cA$ generated by the rule $s\infty\a\infty$.

\begin{lemma}
\label{lem-ord-prop}
\label{lem-le-sub}
Let $a\nf$ be the normal form of $a$ w.r.t. $\a_\cA$.
\begin{lst}{--}
\item $a\eqs b$ iff $a\nf=b\nf$.
\item If $\infty\les a$ or $s^{k+1}a\les a$ then $a\nf=\infty$.
\item If $a\les b$ and $\vphi\les\psi$ then $a\vphi\les b\psi$.
\item If $\vphi\les\psi$ and $U\le V$ then $U\vphi\le V\psi$.
\end{lst}
\end{lemma}

Note that $\infty$-terms are in $\cA$-normal form. The last property
(compatibility of size ordering wrt subtyping) follows from the
restriction $\cV(\vt)=\vide$ in (size).


\section{Decidability of typing}
\label{sec-dec}

In this section, we prove the decidability of type inference and
type-checking for $\infty$-terms under general assumptions that will
be proved in Section \ref{sec-cons}. We begin with some informal
explanations.

How to do type inference? The critical cases are (symb) and (app). In
(symb), a symbol $f$ can be typed by any instance of $\tf$, and two
different instances may be necessary for typing a single term (\eg
$s(sx)$). For type inference, it is therefore necessary to type $f$ by
its most general type, namely a renaming of $\tf$ with fresh
variables, and to instantiate it later when necessary.

Assume now that we want to infer the type of an application $tu$. We
naturally try to infer a type for $t$ and a type for $u$ using
distinct fresh variables. Assume that we get $T$ and $U'$
respectively. Then, $tu$ is typable if there is a size substitution
$\vphi$ and a product type $(x:P)Q$ such that $T\vphi\le(x:P)Q$ and
$U'\vphi\le P$.

After Theorem \ref{thm-sub-s}, checking whether $A\le B$ amounts to
check whether ${A\nf}\le_s {B\nf}$, and checking whether $A\le_s B$
amounts to apply the (prod) rule as much as possible and then to check
that (refl) or (size) holds. Hence, $T\vphi\le{(x:P)Q}$ only if $T\nf$
is a product. Thus, the application $tu$ is typable if
${T\nf}={(x:U)V}$ and there exists $\vphi$ such that
${U'\nf\!\vphi}\le_s {U\vphi}$. Finding $\vphi$ such that $A\vphi\le_s
B\vphi$ amounts to apply the (prod) rule on $A\le_s B$ as much as
possible and then to find $\vphi$ such that (refl) or (size)
holds. So, a subtyping problem can be transformed into a constraint
problem on size variables.

We make this precise by first defining the constraints that can be
generated.


\begin{definition}[Constraints]
{\em Constraint problems} are defined as follows:

\begin{center}
$\cC ::= \bot ~|~ \top ~|~ \cC\et\cC ~|~ a=b ~|~ a\le b$
\end{center}

\noindent
where $a,b\in\cA$, $=$ is commutative, $\et$ is associative and
commutative, $\cC\et\cC=\cC\et\top=\cC$ and $\cC\et\bot=\bot$. A
finite conjunction $\cC_1\et\ldots\et\cC_n$ is identified with $\top$
if $n=0$. A constraint problem is in canonical form if it is neither
of the form $\cC\et\top$, nor of the form $\cC\et\bot$, nor of the
form $\cC\et\cC\et\cD$. In the following, we always assume that
constraint problems are in canonical form. An {\em equality}
(resp. {\em inequality}) {\em problem} is a problem having only
equalities (resp. inequalities). An inequality $\infty\le\alpha$ is
called an {\em $\infty$-inequality}. An inequality $s^p\alpha\le
s^q\b$ is called a {\em linear inequality}. Solutions to constraint
problems are defined as follows:

\begin{lst}{--}
\item $S(\bot)=\vide$,
\item $S(\top)$ is the set of all size substitutions,
\item $S(\cC\et\cD)=S(\cC)\cap S(\cD)$,
\item $S(a=b)=\{\vphi~|~a\vphi=b\vphi\}$,
\item $S(a\le b)=\{\vphi~|~a\vphi\les b\vphi\}$.
\end{lst}

\noindent
Let
$S^\ell(\cC)=\{\vphi~|~\all\alpha,\,{\alpha\vphi\nf}\neq{\infty}\}$ be
the set of {\em linear solutions}.
\end{definition}


We now prove that a subtyping problem can be transformed into
constraints.

\begin{lemma}
\label{lem-inf-cor}
Let $S(U,V)$ be the set of substitutions $\vphi$ such that
$U\vphi\le_s V\vphi$. We have $S(U,V)=S(\cC(U,V))$ where $\cC(U,V)$ is
defined as follows:

\begin{lst}{--}
\item $\cC((x:U)V,(x:U')V')=\cC(U',U)\et\cC(V,V')$,
\item $\cC(C^a\vu,C^b\vv)={a\le b}\et\cE^0(u_1,v_1)\et\ldots\et
\cE^0(u_n,v_n)$ if $|\vu|=|\vv|=n$,
\item $\cC(U,V)=\cE^1(U,V)$ in the other cases,
\end{lst}

\noindent
and $\cE^i(U,V)$ is defined as follows:

\begin{lst}{--}
\item $\cE^i((x\!:\!U)V,(x\!:\!U')V')= \cE^i([x\!:\!U]V,[x\!:\!U']V')=
\cE^i(UV,U'V')\\= \cE^i(U,U')\et\cE^i(V,V')$,
\item $\cE^1(C^a,C^b)={a=b}$,
\item $\cE^0(C^a,C^b)={a=b}\et{\infty\le a}$,
\item $\cE^i(c,c)=\top$ if $c\in\cS\cup\cX\cup\cF\moins\CFB$,
\item $\cE^i(U,V)=\bot$ in the other cases.
\end{lst}
\end{lemma}

\begin{proof}
First, we clearly have $\vphi\in S(\cE^1(U,V))$ iff $U\vphi=V\vphi$,
and $\vphi\in S(\cE^0(U,V))$ iff $U\vphi=V\vphi$ and
$\cV(U\vphi)=\vide$. Thus, $S(U,V)=S(\cC(U,V))$.\qed
\end{proof}


\begin{figure}[ht]
\centering
\caption{Type inference rules\label{fig-tha}}
\begin{tabular}{c}
\\(ax)\quad $\G\tha\st:\B$\quad\quad

(prod)\quad $\cfrac{\G\tha U:\asort_x \quad \G,x:U\tha V:\asort'}
{\G\tha (x:U)V:\asort'}$\\

\\(size)\quad $\G\tha C^\infty:\tC$\quad ($C\in\CFB$)\quad\quad

(symb)\quad $\G\tha f:\tf\ry$\quad ($f\notin\CFB$)\\

\\(var)\quad $\G\tha x:x\G$\quad$(x\!\in\!\dom(\G))$\quad\quad

(abs)\quad $\cfrac{\G\tha U:\asort_x\quad \G,x:U\tha v:V}
{\G\tha{} [x:U]v:(x:U)V}$~$(V\neq\B)$\\

\\(app)\quad $\cfrac{\G\tha t:T \quad \G\tha[\cY\cup\cV(T)] u:U'}
{\G\tha tu:V\vphi\ry\xu}$\quad
\begin{tabular}{c}
(${T\nf}={(x:U)V}$, $\cC={\cC(U'\nf,U)}$,\\
$S(\cC)\neq\vide$, $\vphi=mgs(\cC)$)\\
\end{tabular}\\[3mm]
\end{tabular}
\end{figure}


For renaming symbol types with variables outside some finite set of
already used variables, we assume given a function $\r$ which, to
every finite set $\cY\sle\cZ$, associates an injection $\ry$ from
$\cY$ to $\cZ\moins\cY$. In Figure \ref{fig-tha}, we define a type
inference algorithm $\tha$ parametrized by a finite set $\cY$ of
(already used) variables under the following assumptions:

\begin{enumi}{}
\item It is decidable whether $S(\cC)$ is empty or not.
\item If $S(\cC)\!\neq\!\vide$ then $\cC$ has a most
general solution $mgs(\cC)$.
\item If $S(\cC)\neq\vide$ then $mgs(\cC)$ is computable.
\end{enumi}

It would be interesting to try to give a modular presentation of type
inference by clearly separating constraint generation from constraint
solving, as it is done for ML in \cite{odersky99tapos} for
instance. However, for dealing with dependent types, one at least
needs higher-order pattern unification. Indeed, assume that we have a
constraint generation algorithm which, for a term $t$ and a type
(meta-)variable $X$, computes a set $\cC$ of constraints on $X$ whose
solutions provide valid instances of $X$, \ie valid types for
$t$. Then, in (app), if the constraint generation gives $\cC_1$ for
$t:Y$ and $\cC_2$ for $u:Z$, then it should give something like
$\cC_1\et \cC_2\et (\ex U.\ex V.~ {Y\!=_{\b\eta}\!(x:U)Vx}\et {Z\le
U}\et {X\!=_{\b\eta}\!Vu})$ for $tu:X$.

We now prove the correctness, completeness and minimality of $\tha$,
assuming that symbol types are well sorted ($\th\tf:\asort_f$ for all $f$).


\begin{theorem}[Correctness]
\label{thm-cor-tha}
If $\G$ is a valid $\infty$-environment and $\G\tha t:T$, then $\G\th
t:T$, $t$ is an $\infty$-term and $\cV(T)\cap\cY=\vide$.
\end{theorem}

\begin{proof}
By induction on $\tha$. We only detail the (app) case.

\begin{lst}{}
\item [\bf(app)] By induction hypothesis, $\G\th t:T$,
$\G\th u:U'$ and $t$ and $u$ are $\infty$-terms. Thus, $tu$ is an
$\infty$-term. By Lemma \ref{lem-size-sub}, $\G\th t:T\vphi$ and
$\G\th u:U'\vphi$. Since $T\vphi\nf=(x:U\vphi)V\vphi$, we have
$T\vphi\neq\B$ and $\G\th T\vphi:\asort$. By subject reduction, $\G\th
(x:U\vphi)V\vphi:\asort$. Hence, by (sub), $\G\th
t:(x:U\vphi)V\vphi$. By Lemma \ref{lem-inf-cor}, $S(\cC)=S(U'\nf,U)$
and ${U'\nf\!\vphi}\le_s {U\vphi}$. Since $\G\th U\vphi:\asort'$, by
(sub), $\G\th u:U\vphi$. Therefore, by (app), $\G\th tu:V\vphi\xu$ and
$\G\th tu:V\vphi\ry\xu$ since $\cV(u)=\vide$.\qed
\end{lst}
\end{proof}


\begin{theorem}[Completeness and minimality]
\label{thm-comp-tha}
If $\G$ is an $\infty$-environment, $t$ is an $\infty$-term and $\G\th
t:T$, then there are $T'$ and $\psi$ such that $\G\tha t:T'$ and
$T'\psi\le T$.
\end{theorem}

\begin{proof}
By induction on $\th$. We only detail some cases.

\begin{lst}{}
\item [\bf(symb)] Take $T'=\tf\ry$ and $\psi=\ry^{-1}\vphi$.
\item [\bf(app)] By induction hypothesis, there exist $T$,
$\psi_1$, $U'$ and $\psi_2$ such that $\G\tha t:T$,
$T\psi_1\le(x:U)V$, $\G\tha[\cY\cup\cV(T)] u:U'$ and $U'\psi_2\le
U$. By Lemma \ref{thm-cor-tha}, $\cV(U')\cap\cV(T)=\vide$. Thus,
$\dom(\psi_1)\cap\dom(\psi_2)=\vide$. So, let
$\psi=\psi_1\uplus\psi_2$. By Lemma \ref{thm-sub-s},
${T\nf\!\psi}\le_s {(x:U\nf)V\nf}$. Thus, ${T\nf}={(x:U_1)V_1}$,
${U\nf}\le {U_1\psi}$ and ${V_1\psi}\le {V\nf}$. Since ${U'\psi}\le U$ and
${U\nf}\le {U_1\psi}$, we have ${U'\nf\psi}\le {U_1\psi}$ and, by Lemma
\ref{thm-sub-s}, ${U'\nf\psi}\le_s {U_1\psi}$. Thus, $\psi\in
S(U'\nf,U_1)$. By Lemma \ref{lem-inf-cor}, $S(U'\nf,U_1)=S(\cC)$ with
$\cC=\cC(U'\nf,U_1)$. Thus, $S(\cC)\neq\vide$ and there exists
$\vphi=mgs(\cC)$. Hence, $\G\tha tu:V_1\vphi\ry\t$ where $\t=\xu$. We
are left to prove that there exists $\vphi'$ such that
${V_1\vphi\ry\t\vphi'}\le {V\t}$. Since $\vphi=mgs(\cC)$, there exists
$\psi'$ such that $\vphi\psi'\les\psi$. So, let
$\vphi'=\ry^{-1}\psi'$. Since $\cV(u)=\vide$, $\t$ commutes with size
substitutions. Since ${V_1\psi}\le {V\nf}\le V$, by Lemma
\ref{lem-le-sub}, ${V_1\vphi\ry\t\vphi'}= {V_1\vphi\psi'\t}\le
{V_1\psi\t}\le V\t$.\qed
\end{lst}
\end{proof}


\begin{theorem}[Decidability of type-checking]
Let $\G$ be an $\infty$-environment, $t$ be an $\infty$-term and $T$
be a type such that $\G\th T:\asort$. Then, the problem of knowing
whether there is $\psi$ such that $\G\th t:T\psi$ is decidable.
\end{theorem}

\begin{proof}
The decision procedure consists in (1) trying to compute the type $T'$
such that $\G\tha t:T'$ by taking $\cY=\cV(T)$, and (2) trying to
compute $\psi=mgs(\cC(T',T))$. Every step is decidable.

We prove its correctness. Assume that $\G\tha t:T'$, $\cY=\cV(T)$ and
$\psi=mgs(\cC(T',T))$. Then, $T'\psi\le T\psi$ and, by Theorem
\ref{thm-cor-tha}, $\G\th t:T'$. By Lemma \ref{lem-size-sub}, $\G\th
t:T'\psi$. Thus, by (sub), $\G\th t:T\psi$.

We now prove its completeness. Assume that there is $\psi$ such that
$\G\th t:T\psi$. Let $\cY=\cV(T)$. Since $\G$ is valid and
$\cV(\G)=\vide$, by Theorem \ref{thm-comp-tha}, there are $T'$ and
$\vphi$ such that $\G\tha t:T'$ and $T'\vphi\le T\psi$. This means
that the decision procedure cannot fail ($\psi\uplus\vphi\in
S(T',T)$).\qed
\end{proof}


\section{Solving constraints}
\label{sec-cons}

In this section, we prove that the satisfiability of constraint
problems is decidable, and that a satisfiable problem has a smallest
solution. The proof is organized as follows. First, we introduce
simplification rules for equalities similar to usual unification
procedures (Lemma \ref{lem-simpl-eq}). Second, we introduce
simplification rules for inequalities (Lemma \ref{lem-simpl-ne}). From
that, we can deduce some general result on the form of solutions
(Lemma \ref{lem-form-sol}). We then prove that a conjunction of
inequalities has always a linear solution (Lemma
\ref{lem-max-cost}). Then, by using linear algebra techniques, we
prove that a satisfiable inequality problem has always a smallest
solution (Lemma \ref{lem-poly}). Finally, all these results are
combined in Theorem \ref{thm-dec} for proving the assumptions of
Section \ref{sec-dec}.


Let a {\em state} $\bS$ be $\bot$ or a triplet $\cE|\cE'|\cC$ where
$\cE$ and $\cE'$ are conjunctions of equalities and $\cC$ a
conjunction of inequalities. Let $S(\bot)=\vide$ and $S(\cE|\cE'|\cC)=
S(\cE\et\cE'\et\cC)$ be the solutions of a state. A conjunction of
equalities $\cE$ is in {\em solved form} if it is of the form
$\alpha_1=a_1\et\ldots\et\alpha_n=a_n$ ($n\ge 0$) with the variables
$\alpha_i$ distinct from one another and
$\cV(\va)\cap\{\vec\alpha\}=\vide$. It is identified with the
substitution $\{\vec\alpha\to\va\}$.


\begin{figure}[ht]
\caption{Simplification rules for equalities\label{fig-simpl-eq}}
\centering
$\begin{array}{rr@{~\rightsquigarrow~}l}
\\
(1) & \cE\et sa=sb ~|~ \cE' ~|~ \cC
& \cE\et a=b ~|~ \cE' ~|~ \cC\\

(2) & \cE\et a=a ~|~ \cE' ~|~ \cC
& \cE ~|~ \cE' ~|~ \cC\\

(3) & \cE\et a=s^{k+1}a ~|~ \cE' ~|~ \cC
& \bot\\

(4) & \cE\et\infty=s^{k+1}a ~|~ \cE' ~|~ \cC
& \bot\\

(5) & \cE\et\alpha=a ~|~ \cE' ~|~ \cC
& \cE\{\alpha\!\to\! a\} ~|~ \cE'\{\alpha\!\to\! a\}\et\alpha=a ~|~ \cC\{\alpha\!\to\! a\}
~\mbox{if}~ \alpha\!\notin\!\cV(a)\\[3mm]
\end{array}$
\end{figure}


The simplification rules on equalities given in Figure
\ref{fig-simpl-eq} correspond to the usual simplification rules for
first-order unification \cite{dershowitz90book}, except that
substitutions are propagated into the inequalities.

\begin{lemma}
\label{lem-simpl-eq}
The relation of Figure \ref{fig-simpl-eq} terminates and preserves
solutions: if $\bS_1\rightsquigarrow\bS_2$ then
$S(\bS_1)=S(\bS_2)$. Moreover, any normal form of $\cE|\top|\cC$ is
either $\bot$ or of the form $\top|\cE'|\cC'$ with $\cE'$ in solved
form and $\cV(\cC')\cap\dom(\cE')=\vide$.
\end{lemma}


We now introduce a notion of graphs due to Pratt \cite{pratt77tr} that
allows us to detect the variables that are equivalent to $\infty$. In
the following, we use other standard techniques from graph
combinatorics and linear algebra. Note however that we apply them on
symbolic constraints, while they are generally used on numerical
constraints. What we are looking for is substitutions, not numerical
solutions. In particular, we do not have the constant $0$ in size
expressions (although it could be added without having to change many
things). Yet, for proving that satisfiable problems have most general
solutions, we will use some isomorphism between symbolic solutions and
numerical ones (see Lemma \ref{lem-iso}).

\begin{definition}[Dependency graph]
To a conjunction of linear inequalities $\cC$, we associate a graph
$G_\cC$ on $\cV(\cC)$ as follows. To every constraint $s^p\alpha\le
s^q\b$, we associate the labeled edge
$\alpha\stackrel{p-q}\longrightarrow\b$. The {\em cost} of a path
$\alpha_1\stackrel{p_1}\longrightarrow\ldots
\stackrel{p_k}\longrightarrow\alpha_{k+1}$ is
$\S_{i=1}^k p_i$. A {\em cyclic path} (\ie when
$\alpha_{k+1}=\alpha_1$) is {\em increasing} if its cost is $>0$.
\end{definition}


\begin{figure}[ht]
\caption{Simplification rules for inequalities\label{fig-simpl-ne}}
\centering
$\begin{array}{rr@{~\rightsquigarrow~}ll}
\\
(1) & \cC\et a\le s^k\infty
& \cC\\

(2) & \cC\et\cD
& \cC\et\{\infty\le\alpha~|~\alpha\in\cV(\cD)\}
& \mbox{ if } G_\cD \mbox{ is increasing}\\

(3) & \cC\et s^k\infty\le s^l\alpha
& \cC\{\alpha\to\infty\}\et\infty\le\alpha
& \mbox{ if } \alpha\in\cV(\cC)\\[3mm]
\end{array}$
\end{figure}


A conjunction of inequalities $\cC$ is in {\em reduced form} if it is
of the form $\cC_\infty\et\cC_\ell$ with $\cC_\infty$ a conjunction of
$\infty$-inequalities, $\cC_\ell$ a conjunction of linear inequalities
with no increasing cycle, and
$\cV(\cC_\infty)\cap\cV(\cC_\ell)=\vide$.

\begin{lemma}
\label{lem-simpl-ne}
The relation of Figure \ref{fig-simpl-ne} on inequality problems
terminates and preserves solutions. Moreover, any normal form is in
reduced form.
\end{lemma}


\begin{lemma}
\label{lem-simpl}
If $\cC$ is a conjunction of inequalities then
$S(\cC)\neq\vide$. Moreover, if $\cC$ is a conjunction of
$\infty$-inequalities then $S(\cC)=
\{\vphi~|~\all\alpha\in\cV(\cC),\alpha\vphi\nf=\infty\}$.
\end{lemma}


\begin{lemma}
\label{lem-form-sol}
Assume that $\cE|\top|\cC$ has normal form $\top|\cE'|\cC'$ by the
rules of Figure \ref{fig-simpl-eq}, and $\cC'$ has normal form $\cD$
by the rules of Figure \ref{fig-simpl-ne}. Then,
$S(\cE\et\cC)\neq\vide$, $\cE'=mgs(\cE)$ and every $\vphi\in
S(\cE\et\cC)$ is of the form $\cE'(\up\uplus\psi)$ with $\up\in
S(\cD_\infty)$ and $\psi\in S(\cD_\ell)$.
\end{lemma}

\begin{proof}
The fact that, in this case, $S(\cE)\neq\vide$ and $\cE'=mgs(\cE)$ is
a well known result on unification \cite{dershowitz90book}. Since
$S(\cE\et\cC)= S(\cE'\et\cD)$, $\cV(\cE')\cap\cV(\cD)=\vide$ and
$S(\cD)\neq\vide$, we have $S(\cE\et\cC)\neq\vide$. Furthermore, every
$\vphi\in S(\cE\et\cC)$ is of the form $\cE'\vphi'$ since
$S(\cE'\et\cD)\sle S(\cE')$. Now, since
$\cV(\cD_\infty)\cap\cV(\cD_\ell)=\vide$, $\vphi'=\up\uplus\psi$ with
$\up\in S(\cD_\infty)$ and $\psi\in S(\cD_\ell)$.\qed
\end{proof}

Hence, the solutions of a constraint problem can be obtained from the
solutions of the equalities, which is a simple first-order unification
problem, and from the solutions of the linear inequalities resulting
of the previous simplifications.


In the following, let $\cC$ be a conjunction of $K$ linear
inequalities with no increasing cycle, and $L$ be the biggest label in
absolute value in $G_\cC$. We first prove that $\cC$ has always a
linear solution by using Bellman-Ford's algorithm.

\begin{lemma}
\label{lem-max-cost}
$S^\ell(\cC)\neq\vide$.
\end{lemma}

\begin{proof}
Let $succ(\alpha)=\{\b~|~\alpha\stackrel{p}\longrightarrow\b\in
G_\cC\}$ and $succ^*$ be the reflexive and transitive closure of
$succ$. Choose $\g\in\cZ\moins\cV(\cC)$, a set $R$ of vertices in
$G_\cC$ such that $succ^*(R)$ covers $G_\cC$, and a minimal cost
$q_\b\ge KL$ for every $\b\in R$. Let the cost of a vertex
$\alpha_{k+1}$ along a path
$\alpha_1\stackrel{p_1}\longrightarrow\alpha_2
\stackrel{p_2}\longrightarrow\ldots \alpha_{k+1}$ with $\alpha_1\in R$ be
$q_{\alpha_1}+\S_{i=1}^k p_i$. Now, let $\w_\b$ be the maximal cost
for $\b$ along all the possible paths from a vertex in $R$. We have
$\w_\b\ge 0$ since there is no increasing cycle. Hence, for all edge
$\alpha\stackrel{p}\longrightarrow\b\in G_\cC$, we have
$\w_\alpha+p\le \w_\b$. Thus, the substitution $\vphi=\{\alpha\to
s^{\w_\alpha}\g~|~\alpha\in\cV(\cC)\}\in S^\ell(\cC)$.\qed
\end{proof}


We now prove that any solution has a more general linear
solution. This implies that inequality problems are always satisfiable
and that the satisfiability of a constraint problem only depends on
its equalities.

\begin{lemma}
\label{lem-infty-elim}
If $\vphi\in S(\cC)$ then there exists $\psi\in S^\ell(\cC)$ such that
$\psi\les\vphi$.
\end{lemma}


We now prove that $S^\ell(\cC)$ has a smallest element. To this end,
assume that inequalities are ordered and that $\cV(\cC)=\{\alpha_1$,
\ldots, $\alpha_n\}$. We associate to $\cC$ an adjacency-like matrix
$M=(m_{i,j})$ with $K$ lines and $n$ columns, and a vector $v=(v_i)$
of length $K$ as follows. Assume that the $i$-th inequality of $\cC$
is of the form $s^p\alpha_j\le s^q\alpha_k$. Then, $m_{i,j}=1$,
$m_{i,k}=-1$, $m_{i,l}=0$ if $l\notin\{j,k\}$, and $v_i=q-p$. Let
$P=\{z\in\bQ^n~|~Mz\le v, z\ge 0\}$ and $P'=P\cap\bZ^n$.

To a substitution $\vphi\in S^\ell(\cC)$, we associate the vector
$z^\vphi$ such that $z^\vphi_i$ is the natural number $p$ such that
$\alpha_i\vphi=s^p\b$.

To a vector $z\in P'$, we associate a substitution $\vphi_z$ as
follows. Let $\{G_1,\ldots,G_s\}$ be the connected components of
$G_\cC$. For all $i$, let $c_i$ be the component number to which
$\alpha_i$ belongs. Let $\b_1,\ldots,\b_s$ be variables distinct from
one another and not in $\cV(\cC)$. We define $\alpha_i\vphi_z=
s^{z_i}\b_{c_i}$.

We then study the relations between symbolic and numerical solutions.

\begin{lemma}\hfill
\begin{lst}{--}
\label{lem-sub-vec}
\item If $\vphi\in S^\ell(\cC)$ then $z^\vphi\in P'$. Furthermore, if
$\vphi\les\vphi'$ then $z^\vphi\le z^{\vphi'}$.
\label{lem-vec-sub}
\item If $z\in P'$ then $\vphi_z\in S^\ell(\cC)$. Furthermore, if $z\le z'$
then $\vphi_z\les\vphi_{z'}$.
\label{lem-iso}
\item $z^{\vphi_z}=z$ and $\vphi_{z^\vphi}\qle\vphi$.
\end{lst}
\end{lemma}


Finally, we are left to prove that $P'$ has a smallest element. The
proof uses techniques from linear algebra.

\begin{lemma}
\label{lem-poly}
There is a unique $z^*\in P'$ such that, for all $z\in P'$, $z^*\le z$.
\end{lemma}

An efficient algorithm for computing the smallest solution of a set of
linear inequalities with at most two variables per inequality can be
found in \cite{lueker90siam}. A more efficient algorithm can perhaps
be obtained by taking into account the specificities of our problems.


Gathering all the previous results, we get the decidability.

\begin{theorem}[Decidability]
\label{thm-dec}
Let $\cC$ be a constraint problem. Whether $S(\cC)$ is empty or not
can be decided in polynomial time w.r.t. the size of equalities in
$\cC$. Furthermore, if $S(\cC)\neq\vide$ then $S(\cC)$ has a smallest
solution that is computable in polynomial time w.r.t. the size of
inequalities.
\end{theorem}


\section{Conclusion and related works}

In Section \ref{sec-dec}, we give a general algorithm for type
inference with size annotations based on constraint solving, that does
not depend on the size algebra. For having completeness, we require
satisfiable sets of constraints to have a computable most general
solution. In Section \ref{sec-cons}, we prove that this is the case if
the size algebra is built from the symbols $s$ and $\infty$ which,
although simple, captures usual inductive definitions (since then the
size corresponds to the number of constructors) and much more (see the
introduction and \cite{blanqui04rta}).

A natural extension would be to add the symbol $+$ in the size
algebra, for typing list concatenation in a more precise way for
instance. We think that the techniques used in the present work can
cope with this extension. However, without restrictions on symbol
types, one may get constraints like $1\le\alpha+\b$ and loose the
unicity of the smallest solution. We think that simple and general
restrictions can be found to avoid such constraints to appear. Now, if
symbols like $\times$ are added to the size algebra, then we lose
linearity and need more sophisticated mathematical tools.

The point is that, because we consider dependent types and subtyping,
we are not only interested in satisfiability but also in minimality
and unicity, in order to have completeness of type inference
\cite{chen98thesis}. There exist many works on type inference and
constraint solving. We only mention some that we found more or less
close to ours: Zenger's indexed types \cite{zenger97tcs}, Xi's
Dependent\footnote{By ``dependent'', Xi means constrained types, not
full dependent types.} ML \cite{xi98thesis}, Odersky \mbox{\etal's}
ML with constrained types \cite{odersky99tapos}, Abel's sized types
\cite{abel04ita}, and Barthe \mbox{\etal's} staged types \cite{barthe05tlca}.
We note the following differences:

{\bf Terms.} Except \cite{barthe05tlca}, the previously cited works
consider $\la$-terms {\em \`a la} Curry, \ie without types in
$\la$-abstractions. Instead, we consider $\la$-terms {\em \`a la}
Church, \ie with types in $\la$-abstractions. Note that type inference
with $\la$-terms {\em \`a la} Curry and polymorphic or dependent types
is not decidable. Furthermore, they all consider functions defined by
fixpoint and matching on constructors. Instead, we consider functions
defined by rewrite rules with matching both on constructor and defined
symbols (\eg associativity and distributivity rules).

{\bf Types.} If we disregard constraints attached to types, they
consider simple or polymorphic types, and we consider fully
polymorphic and dependent types. Now, our data type constructors carry
no constraints: constraints only come up from type inference. On the
other hand, the constructors of Zenger's indexed data types must
satisfy polynomial equations, and Xi's index variables can be assigned
boolean propositions that must be satisfiable in some given model (\eg
Presburger arithmetic). Explicit constraints allow a more precise
typing and more function definitions to be accepted. For instance (see
\cite{blanqui04rta}), in order for {\em quicksort} to have type
$list^\alpha\A list^\alpha$, we need the auxiliary {\em pivot}
function to have type $nat^\infty\A list^\alpha\A list^\b\times
list^\g$ with the constraint $\alpha=\b+\g$. And, if {\em quicksort}
has type $list^\infty\A list^\infty$ then a rule like $f~(cons~x~l)\a
g~x~(f~(quicksort~l))$ is rejected since $(quicksort~l)$ cannot be
proved to be smaller than $(cons~x~l)$. The same holds in
\cite{abel04ita,barthe05tlca}.

{\bf Constraints.} In contrast with Xi and Odersky \etal who consider
the constraint system as a parameter, giving DML(C) and HM(X)
respectively, we consider a fixed constraint system, namely the one
introduced in \cite{barthe04mscs}. It is close to the one considered
by Abel whose size algebra does not have $\infty$ but whose types have
explicit bounded quantifications. Inductive types are indeed
interpreted in the same way. We already mentioned also that Zenger
considers polynomial equations. However, his equivalence on types is
defined in such a way that, for instance, $list^\alpha$ is equivalent
to $list^{2\alpha}$, which is not very natural. So, the next step in
our work would be to consider explicit constraints from an abstract
constraint system. By doing so, Odersky \etal get general results on
the completeness of inference. Sulzmann \cite{sulzmann01flops} gets
more general results by switching to a fully constrained-based
approach. In this approach, completeness is achieved if every
constraint can be represented by a type. With term-based inference and
dependent types, which is our case, completeness requires minimality
which is not always possible \cite{chen98thesis}.

{\bf Constraint solving.} In \cite{barthe05tlca}, Barthe \etal
consider system F with ML-like definitions and the same size
annotations. Since they have no dependent type, they only have
inequality constraints. They also use dependancy graphs for
eliminating $\infty$, and give a specific algorithm for finding the
most general solution. But they do not study the relations between
linear constraints and linear programming. So, their algorithm is less
efficient than \cite{lueker90siam}, and cannot be extended to size
annotations like $a+b$, for typing addition or concatenation.

{\bf Inference of size annotations.} As already mentioned in the
introduction, we do not infer size annotations for function symbols
like \cite{chin01hosc,barthe05tlca}. We just check that function
definitions are valid wrt size annotations, and that they preserve
termination. However, finding annotations that satisfy these
conditions can easily be expressed as a constraint problem. Thus, the
techniques used in this paper can certainly be extended for inferring
size annotations too. For instance, if we take $-:nat^\alpha\!\A\!
nat^\b\!\A\! nat^X$, the rules of $-$ given in the introduction are
valid whenever $0\le X$, $\alpha\le X$ and $X\le X$, and the most
general solution of this constraint problem is $X=\alpha$.\\

{\bf Acknowledgments.} I would like to thank very much Miki Hermann,
Hongwei Xi, Christophe Ringeissen and Andreas Abel for their comments
on a previous version of this paper.



\section*{Proofs}
\label{sec-annex}


\subsection{Remark about constraint solving}

One could think of using Comon's work \cite{comon90fcs} but
it is not possible for several reasons:

\begin{lst}{--}
\item We consider two kinds of constraints: equality constraints $a=b$
where $=$ is interpreted by the syntactic equality, and inequality
constraints $a\le b$ where $\le$ is interpreted by the quasi-ordering
$\les$ on size expressions. Instead of large inequalities, Comon
considers strict inequalities $a<b$ where $<$ is interpreted by the
lexicographic path ordering (LPO). Since $\les$ is a quasi-ordering,
we do not have $a\les b\AA a<_\cA b\ou a=b$.

\item Even though one can get rid of $\infty$ symbols in a first step,
thing that we do in Lemmas \ref{lem-form-sol} and \ref{lem-infty-elim},
Comon assumes that there is at least one constant symbol. Indeed, he
studies the ground solutions of a boolean combination of equations and
inequations. However, without $\infty$, we have no ground term. It
does not matter since we do not restrict ourself to ground solutions.
\end{lst}


\subsection{Proof of Lemma \ref{lem-simpl-eq}}

The relation $\rightsquigarrow$ strictly decreases the measure
$(s(\cE),c(\cE))\lex$ where $s(\cE)$ is the number of constraints and
$c(\cE)$ the number of symbols. Its correctness is easily
checked. Now, let $\bS=\cE|\cE'|\cC'$ be a normal form of
$\cE|\top|\cC$. If $\cE\neq\top$ then $\bS$ is reducible. Now, one can
easily check that, if $\cE_1|\cE'_1|\cC_1\rightsquigarrow
\cE_2|\cE'_2|\cC_2$, $\cE'_1$ is in solved form and
$\cV(\cC_1)\cap\dom(\cE'_1)=\vide$, then $\cE'_2$ is in solved form
and $\cV(\cC_2)\cap\dom(\cE'_2)=\vide$. So, $\cE'$ is in solved form
and $\cV(\cC')\cap\dom(\cE')=\vide$.


\subsection{Proof of Lemma \ref{lem-simpl-ne}}

The relation strictly decreases the measure $(c(\cC),v(\cC))\lex$
where $c(\cC)$ is the number of symbols and variables and $v(\cC)$ the
multiset of occurrences of each variable in $\cC$. We now prove the
correctness of these rules. (1) is trivial. (3) follows from Lemma
\ref{lem-ord-prop}. For (2), let
$\cD'=\biget\{\infty\le\alpha~|~\alpha\in\cV(\cD)\}$. We clearly have
$S(\cD')\sle S(\cD)$. Assume that $G_\cD=
\alpha_1\stackrel{p_1}\longrightarrow\ldots
\stackrel{p_k}\longrightarrow\alpha_1$ and $\t\in S(\cD)$.
If $\alpha_i\t\nf=\infty$ then, for all $i$, $\alpha_i\t\nf=\infty$
and $\t\in S(\cD')$. Otherwise, there exist $\g\in\cZ$ and, for all
$i$, $m_i\in\bN$ such that $\alpha_i\t=s^{m_i}\g$, $m_1+p_1\le m_2$,
\ldots, $m_k+p_k\le m_1$. Thus, $\S_{i=1}^k m_i+ \S_{i=1}^k p_i\le
\S_{i=1}^k m_i$. Hence, $\S_{i=1}^k p_i\le 0$ which is not possible
since $G_\cD$ is increasing. Finally, a normal form is clearly in
reduced form.


\subsection{Proof of Lemma \ref{lem-simpl}}

Let $S=\{\vphi~|~\all\alpha\in\cV(\cC),\alpha\vphi\nf=\infty\}$. We
prove that $S\sle S(\cC)$. Let
$\vphi=\{\alpha\to\infty~|~\alpha\in\cV(\cC)\}$ and $a\le b\in\cC$. We
have $a=s^k a'$ and $b=s^l b'$ with $a',b'\in\cZ\cup\{\infty\}$. So,
by Lemma \ref{lem-ord-prop}, $a\vphi=s^k\infty\les b\vphi=s^l\infty$
and $\vphi\in S(\cC)$.

Assume now that $\cC$ is a conjunction of $\infty$-inequalities. Let
$\vphi\in S(\cC)$ and $\alpha\in\cV(\cC)$. Since $\alpha\in\cV(\cC)$,
there exists a constraint $\infty\le\alpha$ in $\cC$. Thus, by Lemma
\ref{lem-ord-prop}, $\alpha\vphi\nf=\infty$ and $\vphi\in S$.


\subsection{Proof of Lemma \ref{lem-infty-elim}}

We can assume w.l.o.g. that $\dom(\vphi)\sle\cV(\cC)$. If, for all
$\alpha\in\cV(\cC)$, $\alpha\vphi\nf=\infty$, then any $\psi\in
S^\ell(\cC)\neq\vide$ works. Otherwise, there exists
$\alpha\in\cV(\cC)$, $\g$ and $p$ such that
$\alpha\vphi=s^p\g$. W.l.o.g., we can assume that $\cC$ has only one
connected component. Let
$D_\ell=\{\alpha\in\dom(\vphi)~|~\alpha\vphi\nf\neq\infty\}$,
$D_\infty=\dom(\vphi)\moins D_\ell$ and $D_\infty'=\{\b\in D_\infty~|~
s^p\alpha\le s^q\b\in\cC\A \alpha\vphi\nf\neq\infty\}$. For every
$\alpha\in D_\ell$, let $\w_\alpha$ be the integer $k$ such that
$\alpha\vphi=s^k\g$. Let $\cC_1=\{s^p\alpha\le s^q\b~|~
\alpha\vphi\nf\neq\infty, \b\vphi\nf\neq\infty\}$,
$\cC_2=\{s^p\alpha\le s^q\b~|~ \alpha\vphi\nf\neq\infty,
\b\vphi\nf=\infty\}$, $\cC_3=\{s^p\alpha\le s^q\b~|~
\alpha\vphi\nf=\infty, \b\vphi\nf=\infty\}$ and
$\cC_3'=\cC_3\uplus\{\b\le\b~|~\b\in D_\infty'\}$. We have
$\cC=\cC_1\uplus\cC_2\uplus\cC_3$. After the proof of Lemma
\ref{lem-max-cost}, by taking $R\sge D_\infty'$ and
$q_\b=max\{KL,\w_\alpha+p-q~|~ s^p\alpha\le s^q\b\in\cC\}$ for every
$\b\in D_\infty'$, there exists $\vphi'\in S^\ell(\cC_3')$. We have
$\dom(\vphi')=\cV(\cC_3')=D_\infty$. Let
$\psi=\vphi|_{D_\ell}\uplus\vphi'$. We clearly have $\psi$ linear
and $\psi\les\vphi$. We now prove that $\psi\in S^\ell(\cC)$. We have
$\psi|_{\cV(\cC_1)}= \vphi|_{\cV(\cC_1)}\in S(\cC_1)$ and
$\psi|_{\cV(\cC_3)}= \vphi'|_{\cV(\cC_3)}\in S(\cC_3)$. Let now
$s^p\alpha\le s^q\b\in\cC_2$. We must check that $s^p\alpha\vphi\le
s^q\b\vphi'$. It follows from the definition of $\vphi'$.


\subsection{Proof of Lemma \ref{lem-sub-vec}}

\begin{lst}{--}

\item Assume that the $i$-th inequality is of the form $s^p\alpha_j\le
s^q\alpha_k$. We must prove that $z^\vphi_j-z^\vphi_k\le q-p$. By
assumption, $s^p\alpha_j\vphi\les s^q\alpha_k\vphi$. Hence,
$p+z^\vphi_j\le q+z^\vphi_k$. The second claim is immediate.

\item Assume that the $i$-th inequality is of the form $s^p\alpha_j\le
s^q\alpha_k$. We must prove that $s^p\alpha_j\vphi_z\les
s^q\alpha_k\vphi_z$, that is, $s^{p+z_j}\b_{c_j}\les
s^{q+z_k}\b_{c_k}$. Since $\alpha_j$ and $\alpha_k$ are connected in
$G_\cC$, $c_j=c_k$. And, by assumption, $z_j-z_k\le q-p$.

\item $z^{\vphi_z}_i$ is the integer $p$ such that $\alpha_i\vphi_z=s^p\b$,
and $\alpha_i\vphi_z=s^{z_i}\b_{c_i}$. Thus, $p=z_i$.

\item $\alpha_i\vphi_{z^\vphi}= s^{z^\vphi_i}\b_{c_i}$, and $z^\vphi_i$ is
the integer $p$ such that $\alpha_i\vphi=s^p\b$. Every variable of a
connected component $c$ is mapped by $\vphi$ to the same variable
$\g_c$. Let $\psi$ be the substitution which associates $\g_c$ to
$\b_c$. We have $\alpha_i\vphi_{z^\vphi}\psi= s^p\b_{c_i}\psi=
s^p\g_{c_i}= \alpha_i\vphi$. Thus, $\vphi_{z^\vphi}\qle\vphi$.

\end{lst}


\subsection{Proof of Lemma \ref{lem-poly}}

Lemma \ref{lem-poly} is Lemma \ref{lem-poly2} (6) below.

See for instance \cite{schrijver86book} for details on {\em
polyhedrons}, \ie sets of the form $\{z\in\bQ^n~|~Mz\le v\}$. Note
that $P=\{z\in\bQ^n~|~M'z\le v'\}$ with $M'=\matc M\\-I\emat$ and
$v'=\matc v\\0\emat$, where $I$ is the identity matrix. We say that a
{\em bit vector} is a vector whose components are in $\{0,1\}$. Given
two vectors $z^a$ and $z^b$, $min\{z^a,z^b\}$ is the vector $z$ such
that $z_i=min\{z^a_i,z^b_i\}$.

\begin{lemma}\hfill
\label{lem-poly2}
\begin{enumi}{}
\item $P$ is {\em pointed}, \ie his lineality space $\{z\!\in\!\bQ^n|M'z=0\}$
has dimension $0$.
\item $P$ is {\em integral}, \ie $P$ is the {\em convex hull} of $P'$.
\item $P$ is infinite.
\item Every minimal proper face of $P$ has for direction a bit vector.
\item If $z^a,z^b\in P$ then $min\{z^a,z^b\}\in P$.
\item There is a unique $z^*\in P'$ such that, for all $z\in P'$, $z^*\le z$.
\end{enumi}
\end{lemma}

\begin{proof}
\begin{enumi}{}
\item If $M'z=0$ then $-Iz=0$ and $z=0$.

\item $P$ is integral since the transpose of $M$ is totally unimodular:
it is a $\{0,\pm 1\}$-matrix with in each column exactly one $+1$ and
one $-1$ (\cite{schrijver86book} p. 274).

\item As any polyhedron, there is a polytope $Q$ such that
$P=Q+char.cone(P)$ (\cite{schrijver86book} p. 88), where
$char.cone(P)=\{z\in\bQ^n~|~M'z\le 0\}$ is the characteristic cone of
$P$. Since every row of $M$ has exactly one $+1$ and one $-1$, the sum
of the columns of $M$ is $0$. Thus, the vector $\mg{1}$ whose
components are all equal to $1$ belongs to $char.cone(P)$ and, either
$P=\vide$ or $P$ is infinite. After Lemma \ref{lem-max-cost},
$S^\ell(\cC)\neq\vide$. Thus, $P$ is infinite.

\item For every minimal proper face $F$ of $P$, there exist a row
submatrix $(L~u)$ of $(M'~v')$ and two rows $(a^i~v'_i)$ and
$(a^j~v'_j)$ of $(M'~v')$ such that $rank(L)=rank(M')-1$ and
$F=\{z\in\bQ^n~|~Lz=u,{}^ta^iz\le v'_i,{}^ta^jz\le v'_j\}$
(\cite{schrijver86book} p. 105). The direction of $F$ is given by
$Ker(L)=\{z\in\bQ^n~|~Lz=0\}$. Let $e^j$ be the unit vector such that
$e^j_j=1$ and $e^j_i=0$ if $i\neq j$. Since $rank(M')=n$,
$rank(L)=n-1$ and there exists $k\le n$ such that $\{Le^j~|~j\neq k\}$
is a family of linearly independent vectors. Thus, $N=\matc
L\\{}^te^k\emat$ is not singular. Let $w= N^{-1}e^k$. If $Lz=0$ then
$Nz=z_ke^k$ and $z=z_kw$. We have $N^{-1}=\cfrac{{}^t com(N)}{det(N)}$
where ${}^t com(N)$ is the transpose matrix of the cofactors of
$N$. Now, one can easily prove that, if every row (or column) of a
matrix $U$ is either $0$, $\pm e^j$ or $e^j-e^k$ with $j\neq k$, then
$det(U)\in\{0,\pm 1\}$. Hence, $det(N)=\pm 1$ and $w$ is a $\{0,\pm
1\}$-vector. The equations satisfied by $z$ in $Lz=0$ are either
$z_i=0$ or $z_i=z_j$. If there is no equation involving $z_i$ then
$Ker(L)=\bQ e^i$ and $w=\pm e^i$. Otherwise, $w\ge 0$ or $w\le
0$. Since $w$ can be replaced by $-w$ w.l.o.g, $w$ can always be
defined as a bit vector.

\item Let $z=min\{z^a,z^b\}$. If $z^a\le z^b$ or $z^b\le z^a$, this
is immediate. Assume now that there are $i\neq j$ such that
$z^a_i<z^b_i$ and $z^a_j>z^b_j$. Since every minimal proper face of
$P$ has for direction a bit vector, we must have $z\in P$.

\item Let $c=min\{\mg{1}z~|~z\in P\}$, $F=\{z\in P~|~\mg{1}z=c\}$,
$z^*\in F$ and $z\in P$. Assume that $z^*\not\le z$. Then,
$z'=min\{z^*,z\}\in P$ and $\mg{1}z'<\mg{1}z^*$, which is not
possible. Thus, $z^*\le z$ and $F=\{z^*\}$. Now, since $P$ is
integral, $z^*\in P'$.\qed
\end{enumi}
\end{proof}


\subsection{Proof of Theorem \ref{thm-dec}}

We can assume that $\cC\neq\bot$. Let $\cC^=$ be the equalities of
$\cC$ and $\cC^\le$ be the inequalities of $\cC$. Compute the normal
form of $\cC^=|\top|\cC^\le$ w.r.t. the rules of Figure
\ref{fig-simpl-eq}. This can be done in polynomial time w.r.t. the
size of equalities. If the normal form is $\bot$ then $S(\cC)=\vide$
and we are done. Otherwise, it is of the form $\top|\cE|\cD$. Let
$\cD_\infty\uplus\cD_\ell$ be the normal form of $\cD$ w.r.t. the
rules of Figure \ref{fig-simpl-ne}. It can be computed in polynomial
time w.r.t. the size of constraints. Let $P=\{z\in\bQ^n~|~M'z\le v'\}$
where $M'$ and $v'$ are the matrix and the vector associated to
$\cD_\ell$. Compute $c=min\{\mg{1}z~|~z\in P\}$ and $z^*\in\{z\in
P~|~\mg{1}z=c\}$. This can be done in polynomial time w.r.t. the size
of constraints since $P$ is integral (see \cite{schrijver86book}
p. 232). Finally, let $mgs(\cC)=\cE(\up\uplus\vphi_{z^*})$ where
$\up\in S(\cD_\infty)$. We prove that this is the smallest solution.

Let $\vphi\in S(\cC)$. By Lemma \ref{lem-form-sol},
$\vphi=\cE(\up'\uplus\vphi')$ where $\up'\in S(\cD_\infty)$ and
$\vphi'\in S(\cD_\ell)$. By Lemma \ref{lem-infty-elim}, there exists
$\psi\in S^\ell(\cD_\ell)$ such that $\psi\qle\vphi'$. By Lemma
\ref{lem-sub-vec}, $z^\psi\in P'$. By Lemma \ref{lem-poly},
$z^*\le z^\psi$. By Lemma \ref{lem-vec-sub}, $\vphi_{z^*}\qle
\vphi_{z^\psi}$. By Lemma \ref{lem-iso},
$\vphi_{z^\psi}\qle\psi$. Thus, $\vphi_{z^*}\qle\vphi'$ and
$mgs(\cC)\qle\vphi$ since $\up\eqs\up'$.

\end{document}